\documentstyle[aps,preprint]{revtex}

\def\prl{{\sl Phys. Rev. Lett.}\ }
\def\PRL{{\sl Phys. Rev. Lett.}\ }
\def\ds{\displaystyle}
\def\pr {{\sl Phys. Rev.}\ }

\def\sss                               {
     \scriptscriptstyle                       }
\def\raise                             {
     {\sss +}                            }
\def\spin                            {
     \sigma                             }
\def\k                               {
     k                            }
\def\kdag{K^\raise}

\def\cspindag                            {
     {\bf d}_{\spin}^{\raise}              }
\def\ckspindag                            {
     {\bf c}_{\k\spin}^{\raise}              }

\begin{document}
\draft

\title{A Variational Ground-State for
 the $\nu=2/3$ Fractional Quantum Hall Regime}
\bigskip
\author{Yigal Meir}
\address{Physics Department, Ben Gurion University,  Beer Sheva 84105,
ISRAEL\\
and \\
Physics Department, University of California,Santa Barbara, CA 93106}
\bigskip
\maketitle
\begin{abstract}
A variational  $\nu=2/3$ state, which unifies the  sharp edge picture
of MacDonald with the soft edge picture of Chang and of Beenakker
is presented and studied in detail. Using
an exact relation between correlation functions of this state and
those of the Laughlin $\nu=1/3$ wavefunction, the correlation functions
of the $\nu=2/3$ state are determined via a classical Monte Carlo calculation,
for systems up to $50$ electrons. It is found that as a function of the
slope of the confining potential there is
a sharp transition of the ground state from one description to the other.
This transition should be observable in tunneling experiments through
quantum dots.
\end{abstract}
\leftline{}
\section{introduction}
The possibility of observing the fractional charge of edge states in the
 fractional quantum Hall regime has generated considerable excitement in recent
years \cite{wen}. While some indications of the fractional charge of the
edge have been reported \cite{leo,simmons},  the situation is far from
resolved. In contrast,  in the integer regime there is a profusion of
information
on the dependence of the edge states on the number of electrons, the
electrostatic potential (gate voltage) and the magnetic field,  obtained
mainly from tunneling measurements through quantum dots \cite{paul}.
This motivated theoretical investigations into the question of tunneling
into a quantum dot in the fractional quantum Hall regime \cite{mac1,jari}.

Recently it was  argued \cite{jari} that tunneling into a $q=\pm e/m$ edge
state will reduce the tunneling amplitude by a factor of $N^{-(m-1)/2}$
 relative to the integer case. Hence, tunneling measurements through a small
  system in the fractional quantum Hall
 regime indeed offer the possibility of directly probing the composition
of the edge structure of the system.
At zero or low magnetic fields
the conductance consists of a series of well separated peaks
\cite{meirav}, each corresponding to an additional electron added to
the system. If the average peak conductance can be studied as a function
 of the number of electrons in the system,  the theory predicts that for
  filling factor $\nu=1/3$,  for example,  the peak amplitudes will fall as
  $1/N$. Since,  however,  tunneling peaks are observed when there is already
   a substantial number of electrons in the system,  the tunneling
   amplitude in this regime may already fall below experimental sensitivity.

   This motivated the suggestion \cite{mac1,jari} that it would be
   advantageous
   to study the $\nu=2/3$ regime,  where the edge structure is
 particularly intriguing,  and
several theories have been proposed to describe the edge states.
 One picture, due to MacDonald \cite{macdonald}, is
based upon a wavefunction proposed by Girvin \cite{girvin}, which, due to
particle-hole symmetry, consists of droplet of
holes in the $\nu=1/3$ Laughlin-state\cite{laughlin}
 embedded in a droplet of electrons
in the $\nu=1$ state.
 This wavefunction has indeed been shown to be an excellent description
of the exact ground-state for a system with a small number of electrons
under various boundary conditions
\cite{mac1,jari,yoshi}. For example, Greiter\cite{yoshi} quotes an overlap
of $0.9990$ between the exact ground state and the Girvin wavefunction for
a system of $8$ electrons in spherical geometry.
This $\nu=2/3$ wavefunction supports two different edges\cite{macdonald},
one at the edge of the hole droplet (of charge $q=-e/3$),
and the other at edge of the $\nu=1$ electron droplet (of charge $q=e$).
On the other hand, a very different edge structure was suggested
by Chang and by Beenakker \cite{beenfrac}, and elaborated on by Chklovskii et
al. \cite{chklovskii} in a more general context. They
 argued  that for a smooth enough potential
an incompressible  $\nu=1/3$
state will nucleate near the edges of the system, leading
again to two edge branches, but this time of charge $q=e/3$ each
 \cite{more}. In the first scenario, where a single edge state
carries a fractional
charge, one would expect that half of the tunneling
peaks will be suppressed, giving
a clear signature of the composition of the edge states. In the second
scenario, where both edges carry a fractional charge, all the peaks would
be suppressed, resulting in a very low conductance signal.

In this work we study quantitatively the nature of the ground state and
the corresponding edge states in the $\nu=2/3$ regime,  in order
 to understand which of the pictures is relevant experimentally.
  Generalizing the Girvin
wavefunction to incorporate the possibility of a $\nu=1/3$ strip near the
edge of the sample, the correlation functions in this
generalized state are expressed exactly in terms of correlation functions
calculated in the $\nu=1/3$
Laughlin wavefunction. Using the mapping onto
a classical one-component two-dimensional plasma \cite{laughlin} we
calculate those  correlation functions
using classical Monte Carlo \cite{metro} for up to $50$ electrons.
The resulting $\nu=2/3$ correlation functions enable us to calculate the
energy of the state for arbitrary electron-electron interactions and
confining potential. We find that as a function of the slope of
the confining potential,
the ground state makes a sharp transition from
the Girvin-MacDonald form  to the Chang-Beenakker form.
This calculation suggests that for heterostructures where the gates are not
too far from the two-dimensional electron gas, the suppression of
half of the peaks, in the first scenario above, should
be observable. In addition, information about the actual distance
between the two edges, which is a crucial ingredient of recent
edge state theories \cite{wen}, is obtained.
The main results of this work have been published in a short communication
\cite{myfqh}. The results have also been verified by
composite-fermion calculations \cite{brey}.

The rest of the paper is organized as follows. In the following section we
introduce the variational wavefunction and describe its construction.
in Sec.III the  correlation functions for this wavefunction
are expressed in terms of the correlation functions for the $\nu=1$ and the
 $\nu=1/3$ systems,  with more details in the appendix. Sec.IV describes
 the numerical evaluations of the $\nu=1/3$ correlation function,  while
 Sec.V contains the main results for the  $\nu=2/3$  state. Sec.VI summarizes
  and concludes.

\section{The Variational Wavefunction}

The ground state of $N$ electrons in a radially symmetric
system in the $\nu=1/3$ fractional quantum
Hall regime can be approximated very well by the Laughlin wavefunction
\cite{laughlin},
\begin{equation}
\Psi^{(\sss 1/3)}\left(z_1,...,z_N\right) =
\prod_{i<j}\,(z_i-z_j)^3\ e^{-\sum_i|z_i|^2/4} \ \
 \hat{=}\sum_{\{i_1,\cdots,i_N\}} C^{\sss (N)}_{\{i_1,\cdots,i_N\}}\
 {\bf a}^{\raise}_{i_1}\, \cdots\,{\bf a}^{\raise}_{i_N}\ |\,0> \ \ ,
\label{eq:laughlin}
\end{equation}
where $z_i$ denotes the complex coordinates of the $i$-th particle in the
plane,
and all lengths are expressed in units of the magnetic length,
$\ell_H\equiv \sqrt{\hbar c/eH}$. $\hat{=}$ denotes a second-quantization
representation, where ${\bf a}^{\raise}_{n}$ creates an electron in
 a first Landau-level state of angular momentum $n$, described by the
single-particle wavefunction $\phi_n(z)=z^n\exp(-|z|^2/4)/\sqrt(2\pi 2^n n!)$.
The sum is over all permutations of $N$ distinct integers which sum up
to the total angular momentum
$3N(N-1)/2$, and the $C^{(N)}_{\{i_1,\cdots,i_N\}}$ can, in principle,
be obtained by expanding the first product.

The particle-hole symmetry-based wavefunction, introduced
by Girvin to describe the $\nu=2/3$ state \cite{girvin}, consists of $N_h$
holes in the $\nu=1/3$ state embedded in a first Landau level state of
$N+N_h$ electrons,  and can
be written in a second-quantized form as
\begin{equation}
\Psi^{(\sss 2/3;N_h)}\left(z_1,...,z_N\right)  \
\hat{=}\sum_{\{i_1,\cdots,i_{\sss N_h}\}} C^{\sss (N_h)}_{\{i_1,\cdots,i_{\sss
N_h}\}}\
 {\bf a}_{i_1}\, \cdots\,{\bf a}_{i_{\sss N_h}}\
{\bf a}^{\raise}_1\, \cdots\,{\bf a}^{\raise}_{\sss N+N_h}\ |\,0> \ \ .
\label{eq:girvin}
\end{equation}
The yet undetermined number of holes, $N_h$,
must be chosen to minimize the energy.

In order to allow for the possibility of a $\nu=1/3$ state nucleating
along the edge of the sample, we start with
the Laughlin wavefunction with an inside hole of size $L$\cite{laughlin},
\begin{equation}
\Psi^{(\sss L;1/3)}\left(z_1,...,z_N\right) =
\prod_{i} z_i^{\sss L} \prod_{i<j}\,(z_i-z_j)^3\ e^{-\sum_i|z_i|^2/4} \ \
 \hat{=}\sum_{\{i_1,\cdots,i_{\sss N}\}} D^{\sss (N;L)}_{\{i_1,\cdots,i_N\}}\
 {\bf a}^{\raise}_{i_1+{\sss L}}\,
\cdots\,{\bf a}^{\raise}_{i_{\sss N}+{\sss L}}\ |\,0> \ \ ,
\label{eq:shifted}
\end{equation}
where the sum is over the same sets as in (\ref{eq:laughlin}),
and $D^{\sss (N;L)}$
can again, in principle, be evaluated by expanding the products.
(Clearly $D^{\sss (N;0)}=C^{\sss (N)}$ for each set of integers.)
With the above  wavefunction describing a strip of electrons with $\nu=1/3$
correlations along the edge of the sample, we write a generalized
$\nu=2/3$ state,
\begin{eqnarray}
\Psi^{(\sss 2/3;N_h,L,N_2)}\left(z_1,...,z_N\right)&\nonumber \\
\hat{=} \ds{\sum_{\{i_1,\cdots,i_{_{\sss N_h}}\}}
\sum_{\{j_1,\cdots,j_{_{\sss N_2}}\}}}
& C^{\sss (N_h)}_{\{i_1,\cdots,i_{_{\sss N_h}}\}}\
D^{\sss (N_2;L)}_{\{j_1,\cdots,j_{_{\sss N_2}}\}}\
 {\bf a}_{i_1}\, \cdots\,{\bf a}_{i_{\sss N_h}}\
{\bf a}^{\raise}_1\, \cdots\,{\bf a}^{\raise}_{\sss N_1+N_h}\
{\bf a}^{\raise}_{j_1+{\sss L}}\,
\cdots\,{\bf a}^{\raise}_{j_{_{\sss N_2}}\!\!+{\sss L}}\ |\,0> \ \ .
\label{eq:tt}
\end{eqnarray}
This wavefunctionis schematically depicted in Fig.1. It depends on
three integer parameters. Out of the $N$ electrons,  $N_1$ are described by
the Girvin wavefunction (\ref{eq:girvin}), with
$N_h$ holes.
The remaining $N_2=N-N_1$ electrons nucleate into
a $\nu=1/3$ strip along the edge of the sample, with minimal
angular momentum $L$ ($L>N_1+N_h$).

\section{The Correlation Functions}

Our task now is to find the set of parameters $N_h, N_1$,  and $L$,
 that minimizes the energy of an $N$-particle state
for a given  confining potential and interactions.
 To
this end we need to calculate the one-particle and two-particle
correlation functions for this state,
\begin{eqnarray}
\rho_1^{(N)}(r) &=& \int dz_1\cdots dz_N\
|\Psi^{(N)}\left(z_1,...,z_N\right)|^2 \delta(|z_1|^2-r^2) \nonumber \\
\rho_2^{(N)}(z_1,z_2) &=& \int dz_3\cdots dz_N\
|\Psi^{(N)}\left(z_1,...,z_N\right)|^2  .
\label{eq:correlations}
\end{eqnarray}
The normalization is chosen such that $\rho_1(r)$ is normalized to $N$,
and
$\rho_2(r)\equiv\int dz_1\int
dz_2\,\rho_2^{(N)}(z_1,z_2)\delta(|z_1-z_2|^2-r^2)$
 is normalized to $N-1$ \cite{difference}. The energy
is easily obtained from these correlation functions,
\begin{equation}
E = 2\pi\int r dr \left[ V(r) \rho_1^{(N)}(r) + N U(r) \rho_2^{(N)}(r) \right],
\end{equation}
where $V(r)$ and $U(r)$ are the potential energy and the interaction,
 respectively.\par
In principle, of course, if one
can obtain the coefficients $C^{\sss (N_h)}_{\{i_1,\cdots,i_{_{\sss N_h}}\}}$
and $D^{\sss (N_2;L)}_{\{j_1,\cdots,j_{_{\sss N_2}}\}}$, all correlation
functions
for the $\nu=2/3$ state (Eq.(\ref{eq:tt}))
chould be readily evaluated. This, however, can only be achieved
for a system of a very small number ($N\leq6$) of particles \cite{jari,expand}.
Correlations functions for the $\nu=1/3$
Laughlin-like wavefunction(\ref{eq:shifted}), on the other hand,
 can be straightforwardly calculated for a large number of particles, using
a mapping into a classical statistical problem \cite{laughlin}.
Such a calculation, however, is not possible for the Girvin wavefunction
(\ref{eq:girvin}), or the generalized wavefunction (\ref{eq:tt}), due to
a resulting sign problem.

The most important step in this work is expressing the
correlation functions for the generalized $\nu=2/3$ wavefunction
in terms of correlation functions for the Laughlin-like wavefunctions
for $\nu=1/3$. The main steps in this mapping are described in the
appendix. The results are
\begin{eqnarray}
\rho_1^{(\sss 2/3;N;N_h,L,N_2)}(r) &=& \rho_1^{(\sss 1;N_1)}(r) -
\rho_1^{(\sss 1/3;N_h)}(r) + \rho_1^{(\sss 1/3;N_2;L)}(r)
\nonumber \\
N \rho_2^{(\sss 2/3;N;N_h,L,N_2)}(z_1,z_2) &= &
 N_1 \rho_2^{(\sss 1;N_1)}(z_1,z_2)
+ N_h \rho_2^{(\sss 1/3;N_h)}(z_1,z_2) + N_2 \rho_2^{(\sss 1/3;N_2;L)}(z_1,z_2)
\nonumber \\
-&&\!\!\!\!\!\!\!\!\!\!\!\!\rho_1^{(\sss 1;N_1)}(r_1)\rho_1^{(\sss
1/3;N_h)}(r_2) -
 \rho_1^{(\sss 1;N_1)}(r_2)\rho_1^{(\sss 1/3;N_h)}(r_1) +
\rho_1^{(\sss 1;N_1)}(r_1)\rho_1^{(\sss 1/3;N_2;L)}(r_2) \nonumber \\
+&&
\!\!\!\!\!\!\!\!\!\!\!\!\rho_1^{(\sss 1;N_1)}(r_2)\rho_1^{(\sss
1/3;N_2;L)}(r_1)
-\rho_1^{(\sss 1/3;N_h)}(r_1)\rho_1^{(\sss 1/3;N_2;L)}(r_2) -
\rho_1^{(\sss 1/3;N_h)}(r_2) \rho_1^{(\sss 1/3;N_2;L)}(r_1)
\nonumber \\
+& \ \ 2 Re& \left[\ds{\sum_{i=0}^{3(N_h-1)}}\ \
\ds{\sum_{j=0}^{N_1-1}}<\!n_i\!>_{\!\!_{1/3}}^{\!\!^{(N_h)}}
 \phi_i^*(z_1) \phi_i(z_2)
\phi_j^*(z_2) \phi_j(z_1)\right]
\nonumber \\
-& \ \ 2 Re& \left[\ds{\sum_{i=L}^{L+3(N_2-1)}}\ \
\ds{\sum_{j=0}^{N_1-1}}<\!n_i\!>_{\!\!_{1/3}}^{\!\!^{(N_2;L)}}
 \phi_i^*(z_1) \phi_i(z_2)
\phi_j^*(z_2) \phi_j(z_1)\right]
\ \ \ \ \ \ \ \ \ \ \ \ \ \ \  \ \ \ \ \ \ \ \ \ \
\nonumber \\
+& \ \ 2 Re &\left[\ds{\sum_{i=L}^{L+3(N_2-1)}}\ \
\ds{\sum_{j=0}^{3(N_h-1)}}<\!n_i\!>_{\!\!_{1/3}}^{\!\!^{(N_2;L)}}
<\!n_j\!>_{\!\!_{1/3}}^{\!\!^{(N_h)}}
 \phi_i^*(z_1) \phi_i(z_2)
\phi_j^*(z_2) \phi_j(z_1)\right]  ,
\label{eq:mapping}
\end{eqnarray}
with $r_i=|z_i|$.
The single-particle distribution function $\rho_1$, is simply
expressed as the sum of the three distribution functions of the $N_1$ electrons
in the $\nu=1$ state ($\rho_1^{(1;N-N_2)}$),  that of the $N_2$ electrons
in the strip of $\nu=1/3$ state ($\rho_1^{(1/3;N_2;L)}$), and (minus) that of
 the $N_h$ holes in the $\nu=1/3$ state,
($\rho_1^{(1/3;N_h)}$). The two-particle correlation function, $\rho_2$,
 is far more complicated. Nevertheless, the various terms contributing
to the resulting interaction energy have straightforward interpretations.
The first three terms describe the contribution to the interaction energy
from interactions within the three different components. The next six
terms describe the direct (Hartree) interactions between the three components,
the $N_h$ holes in the $\nu=1/3$ state,  the $N_1+N_h$ electrons in the
 $\nu=1$ state, and the $N_2$ electrons in the $\nu=1/3$ strip.
The last three nontrivial terms correspond to the exchange and correlation
interactions between the different components.

\section{The Correlation Function for the $\nu=1/3$ State}

Eq.(\ref{eq:mapping}) enables us to express the one- and two-particle
correlation functions for the $\nu=2/3$
state in terms of quantities evaluated for the $\nu=1/3$
states (Eq.\ref{eq:laughlin} and Eq.\ref{eq:shifted}). We calculate
the $\nu=1/3$  correlation functions
using classical Monte Carlo   method \cite{metro,datta},  based on the
mapping of expectation values in the Laughlin state into statistical
correlation functions for a two-dimensional classical plasma  \cite{laughlin},
which for the wavefunction (\ref{eq:shifted}) takes the form

\begin{equation}
|\Psi^{(\sss L;1/3)}\left(z_1,...,z_N\right) |^2 = \exp \left[-\sum_i|z_i|^2/2
+
3\sum_{i<j} \log(|z_i-z_j|^2) +
 L\sum_i\log(|z_i|^2) \right] .
\label{eq:plasma}
\end{equation}

$L=0$ corresponds to the special case Eq.(\ref{eq:laughlin}).
The term in the brackets can now be considered as the Hamiltonian
for a classical system, for which Monte Carlo calculations can be applied.
Fig. 2 depicts the one and two-particle correlation functions for
 $20$ and $50$ particles. Similar results can be easily obtained for
 larger systems. As the number of particles increases,  the edge
 structure in the single-particle density shifts to larger distances,
 without a significant change in the bulk density,  while the two-particle
 correlation function reaches a similar maximum,  but decays on a longer
 length scale.

The main numerical problem in the present work is not the calculation of
 the correlation functions for large number of particles,  but the increasing
  number of possible parameter sets that needs to be considered. Accordingly
   we limit ourselves to systems with up to $50$ electrons.

 In addition we have to calculate
 $<\!n_i\!>_{1/3}$, the average occupation of the $i$-th state. This
 quantity is somewhat more complicated to calculate, and so far it has
  been calculated with only a partial success \cite{mac3}. The main
  difficulty is that the expression of the density in terms of
   the occupation numbers,
\begin{equation}
\rho_1^{(\sss 1/3;N)}(r) = \sum_{i=0}^{3(N-1)} <\!n_i\!>_{\!\!_{1/3}}
  |\phi_i(r)|^2 ,
\label{eq:occupations}
\end{equation}
 involves nonorthogonal functions,  $|\phi_i(r)|^2$. The significant
 overlap between them makes impossible the extraction of the occupations
 from the density profile \cite{mac3}.
 In this work we use a novel idea to calculate
 the occupations using correlation functions for an $N-1$-state,
 \begin{eqnarray}
 \rho_1^{(\sss 1/3;N)}(r) &=&
 \int dz_1\cdots dz_N\ |\Psi^{(\sss 1/3;N)}\left(z_1,...,z_N\right)|^2 \
 \delta(|z_1|^2-r^2) \nonumber \\
  &=& \int dz_1\cdots dz_N \ \prod_{i=2}^N |z_1-z_i|^6\,
 e^{-|z_1|^2/2} \ |\Psi^{(\sss 1/3;N-1)}\left(z_2,...,z_N\right)|^2 \
 \delta(|z_1|^2-r^2)
  \nonumber \\
 &=&
 \sum_{i=0}^{3(N-1)}\left[ \int dz_2\cdots dz_N \ F_i(z_2,\cdots,z_n)\
 |\Psi^{(\sss 1/3;N-1)}\left(z_2,\cdots,z_N\right)|^2 \right] |\phi_i(r)|^2 ,
 \label{nm}
 \end{eqnarray}
 where $F_i$ can be easily obtained by expanding the Laughlin wavefunction.
 Comparing Eqs.(\ref{eq:occupations}) and (\ref{nm}) shows that
 the occupations for the $N$-electron system,  $<\!n_i\!>_{\sss 1/3}$,   can be
 directly evaluated by numerically calculating the averages of
   the functions $F_i$
 in the $N-1$-particle system. In Fig. 3 we show a comparison
 between the one-particle distribution function
 deduced from the occupations such calculated and the independently calculated
 distribution function for $20$ particles. An excellent agreement is observed.
  In addition,  the occupations near the edge
  were found to obey various exact relations
  obtained by directly expanding the Laughlin wavefunction\cite{mac3}.

\section{Results for $\nu=2/3$}

Having obtained the correlation function for the generalized
$\nu=2/3$ state (\ref{eq:tt}), its energy can easily be evaluated for
any choice of interactions and confining potential. In order to be as
close to the experimental situation as possible, we present results for
Coulomb interactions, $U(r)=e/\epsilon r$.  Rather similar results
have been found for other types of interactions, such as screened
or logarithmic ones.

As all the wavefunctions have uniform bulk density, they differ only by
 their structure near the edge,  or,  alternatively,  by the occupations
 of the single-electron states near the Fermi energy. Accordingly, as
 indeed verified numerically, the only relevant ingredient of the confining
 potential is its slope at the Fermi energy. The detailed results will
 be presented for a linear confining potential,  of slope $S/d$,
 rising from zero to $S$ over the range  $(r_0-d/2,r_0+d/2)$.
The position of the
 midheight of the potential step, $r_0$, is fixed so the filling factor
is $2/3$.
As discussed below,
the physically relevant parameter will be the slope of the potential,
$S/d$.
$d$ is determined
experimentally by the distance of the gates from the two-dimensional
electron gas, while $S$ is determined by the amount of voltage applied
 to the gate, as seen by the electrons in the 2d gas. For typical
Ga-As samples, the gates are 120-200nm from the 2d gas, which corresponds
to 8-12 magnetic lengths.
The interaction energy $e/\epsilon \ell_H$ is
typically 5 meV, while the boundary potential seen by the electrons
is tens of meV \cite{frank}. Here we will express all energies in units
of $e/\epsilon \ell_H$. The calculations were done for
up to $50$ electrons, which is a typical
number in an experimental quantum dot\cite{meirav}.

In Fig.4a we plot the number of holes, $N_h$, which minimizes the energy
for a step potential ($d=0$), for two values of $S=3$ and $S=5$. For
a step potential, the ground state usually involves $N_2=0$ electrons in the
$\nu=1/3$ strip, so it is of the Girvin type (\ref{eq:girvin}).
The number of holes in the ground state is determined by the competition
between the two contributions to the energy: the larger the number of
holes, the more uniform the density, and the lower the
interaction energy. On the other hand, the larger the number of holes,
the more angular-momentum states are occupied,  and
 the larger the potential energy.
Thus,
as the potential becomes softer,
the number of holes may increase and a strip of electrons in the $\nu=1/3$
state may form near the edge.

As can be seen from the figure, the number of holes in the ground state
scales as $N/2$, which shows the region of density different from $\nu=2/3$
is independent of $N$, namely an edge effect. From Fig.4a one can obtain
the physical distance between the edges for large enough system, as a function
of the confining potential. We find that this distance changes from
$\sim 1.5\ell_H$ to $\sim 2.5\ell_H$ when $S$ changes from $3$ to $10$.
Thus, unlike the case for slowly varying confining potential\cite{chklovskii}
 one cannot consider those edges as isolated from each other, and
any effective theory should include interactions and mixing of those
 states \cite{fisher}.

Since the number of holes is an integer, it
 will change, on average, every other time an electron is added to
the system. This is the source of the prediction \cite{jari} that half
of the peaks for  tunneling into a $\nu=2/3$ droplet
will be suppressed. As the present calculation cannot
produce the tunneling amplitudes exactly, we estimate them by
their upper limit,
the average occupation of the angular momentum state
the electron tunnels to.
In the inset of Fig.4a we plot this occupation as a function of $N$.
The suppression of more than
half of the peaks is clearly observed, with the right
power-law dependence on the electron number. Interestingly, the
calculation suggests that sometimes the ground states of $N$ and $N+1$
electrons differ by two holes. It remains to be seen if this is a real
effect, which will result in a more dramatic reduction of the peak
amplitude.

In any real system the potential will rise over a finite length scale, $d$.
We have studied the nature of the ground state as a function of $d$,
and we found that for a given electron number $N$, and potential height $S$,
there will be a transition from the ground state being of
the Girvin type (\ref{eq:girvin}) to a state which includes electrons
nucleating at the edge of the sample in the $\nu=1/3$ state. By varying
$S$ it is found the transition occurs at the same ratio of $S/d$, namely
at a given slope of the confining potential.
For example,
 for $30$
electrons the ground state evolves smoothly from
$N_h=9$ and $N_2=0$, for $d=0$, to  $N_h=5$ and $N_2=0$ for $S/d\simeq1$,
and then it changes abruptly
 to $N_h=15$, $N_2=2$ and $L=N_1+N_h+1$
 (Fig. 4b). Thus the two edges,
of the electron droplet and the hole droplet  suddenly merge, and
a $\nu=1/3$ strip forms, signaling a transition from the
Girvin-MacDonald picture to the
picture presented
by Chang and by Beenakker \cite{beenfrac} and Chklovskii et al.
\cite{chklovskii}.
This
strip moves away from the edge of the electron droplet ($L=N_1+N_h+1$)
as $S/d$ decreases. For example, for $S/d=0.6$ the ground state corresponds to
$N_h=15$, $N_2=2$ and $L-(N_1+N_h)=20$ (Fig. 2b).
For $40$ electrons  one can actually see
two transitions. For $S/d\simeq1$ the ground state changes from $N_h=12$
and $N_2=0$ to   $N_h=19$ and $N_2=0$, namely it is still described by
Eq.(\ref{eq:girvin}), but the two edges have merged, while for
$S/d\simeq1.4$ nucleation first occurs with $N_h=18$ and $N_2=5$.
This intermediate regime where the two edges merged may suggest
a possible description in terms of a single $\nu=2/3$ edge \cite{fisher}.

Similar transitions have been observed for other forms of confining
potentials and electron-electron interactions.
As the slope of the potential in
experimental systems\cite{meirav} is if the order of
$1.2-3$ $e^2/\epsilon \ell_H^2$ \cite{frank},
we predict that while the experiment still being in the smooth side
of the transition,  the suppression of half of the tunneling peaks should
be observable in quantum dots in present high mobility structures.
 The closer the gates to the 2d gas, the better the chances
 of seeing that effect.  In addition,  it is predicted
that as a function of the voltage applied to the gates, (which
changes the slope of the effective potential), the tunneling
peak structure will change abruptly as this transition occurs. For high
voltages half of the peaks appearing in the $\nu=1$ regime will be
suppressed in the $\nu=2/3$ regime, while
for lower voltages, as extensive tunneling into the $\nu=1/3$ state
will occur, most or all of the peaks will be suppressed.

\section{Summary and Conclusions}
In conclusion, using an exact expression for the generalized $\nu=2/3$ state
 correlation
functions in terms of the $\nu=1/3$ and the $\nu=1$ ones,
 we have been able to study
quantitatively systems of relatively large number of electrons ($N\leq 50$).
Consequently we predict a transition in the nature of the
ground state of the system as a function of the slope of the
confining potential and discuss its experimental
manifestation.

In this work it was assumed that all electrons are  spin polarized due to
the strong magnetic field. There is experimental and numerical evidence
that there could be a density and magnetic field regime,  where the
$\nu=2/3$ state can involve both spin directions. The transition between
this latter state and the spin-polarized one can be also explored using similar
methods to the one described in this work.

It is hoped that this work will stimulate more experiments
in this direction.

\section{Acknowledgments}
I thank M. P. A. Fisher, W. Kohn and X.-G. Wen for useful
discussions.
Work in Santa Barbara
 was supported by NSF grant no. NSF-DMR90-01502  and by the
NSF Science and Technology Center for Quantized Electronic Structures,
Grant no. DMR 91-20007. The numerical calculations in this work have
been performed on the San Diego Supercomputer CRAY-YMP. Additional
calculations have been performed using resources of the Cornell Theory
Center.

\appendix
\section*{}
Here we derive the relations between the correlation function for
the $\nu=2/3$ wavefunction (Eq.\ref{eq:tt}) and those for the $\nu=1$
and $\nu=1/3$ wavefunctions. For simplicity we detail the derivation
for the Girvin wavefunction (\ref{eq:girvin}). The generalization to
the wavefunction (\ref{eq:tt}) is straightforward.

We start from the definitions (\ref{eq:correlations}),  where we expand
 $\Psi$ in terms of Slater determinants (e.g. Eqs.(\ref{eq:laughlin}) or
 (\ref{eq:girvin})) . To calculate
$\rho_1$,  one can integrate over $z_2, \cdots,  z_N$. The result will be
nonzero only if $N-1$ of the quantum numbers of the the two Slater determinants
involved are the same. Since the total angular momentum is fixed that
means that the last quantum number is also identical. This leads to an equation
similar  to (\ref{eq:occupations}),  with $<n_i>$ the average occupations
in this particular state. Since,  by definition,
the occupations of the Girvin wavefunction
(\ref{eq:girvin}) are trivially related to those of the full Landau
level and those of the Laughlin wavefunction,
$n_i^{(2/3)}=n_i^{(1)}- n_i^{(1/3)}$,
the first of Eqs.(\ref{eq:mapping})  immediately follows.

The derivation of the second equation is more complicated. To calculate
$\rho_2$,  one integrates over $z_3, \cdots,  z_N$.
Again, because of fixed total angular momentum, the sum of the remaining
two quantum numbers should be the same for the two Slater determinants.
Going over the possibilities,  one finds
\begin{eqnarray}
\rho(z_1, z_2)&=& \sum_{n_1, n_2} <n_1 n_2> \left[ |\phi_{n_1}(z_1)|^2
|\phi_{n_2}(z_2)|^2 -
\phi_{n_1}^*(z_1)\phi_{n_2}^*(z_2)\phi_{n_2}(z_1)\phi_{n_1}(z_2) \right]
\nonumber\\
&+& \sum_{\{n_1, n_2\}\ne\{n_3, n_4\}}
<{\bf a}^{\raise}_{n_1}{\bf a}^{\raise}_{n_2}{\bf a}_{n_3}{\bf a}_{n_4}>
\phi_{n_1}^*(z_1)\phi_{n_2}^*(z_2)\phi_{n_2}(z_1)\phi_{n_1}(z_2) .
\label{eq:appendix1}
\end{eqnarray}
This equation is general. Next we need to relate the expectation values
in the Girvin wavefunction to the expectation values
in the full Landau
level and in the Laughlin wavefunction. A straightforward calculation
gives
\begin{eqnarray}
<n_k n_m>_{2/3} = <n_k n_m>_{1} &-&
<n_k>_{1/3} <n_m>_1 - <n_m>_{1/3} <n_k>_1
 +  <n_k n_m>_{1/3}\nonumber\\
<{\bf a}^{\raise}_{n_1}{\bf a}^{\raise}_{n_2}{\bf a}_{n_3}{\bf a}_{n_4}>_{2/3}
&=&
<{\bf a}^{\raise}_{n_3}{\bf a}^{\raise}_{n_4}{\bf a}_{n_1}{\bf a}_{n_2}>_{1/3}
,
\label{eq:appendix2}
\end{eqnarray}
where the second equation applies only for $\{n_1, n_2\}\ne\{n_3, n_4\}$.
Combining these equations and trivially generalizing to the
wavefunction (\ref{eq:tt}),  give rise to the relations (\ref{eq:mapping}).

\newpage
\vskip 8truecm
\leftline{\sl Figure Captions}

1. Schematic representation of the generalized $\nu=2/3$ state (\ref{eq:tt}).
 Out
 of the $N$ electrons,
$N_1$ are described by the Girvin wavefunction, consisting
 of $N_h$ holes in the $\nu=1/3$ state
in the background of $N_1+N_h$ particles in
 the $\nu=1$ state (\ref{eq:girvin}).
The remaining $N_2$ electrons nucleate into a $\nu=1/3$ strip
 along the edge of the sample.

2. Correlation functions for the $\nu=1/3$ obtained from classical Monte
 Carlo calculations.

3. Comparison of the density profile obtained from classical Monte
 Carlo calculations for $20$ particles (circles),  and the  density profile
 derived from the occupations calculated using  Eqs.(\ref{eq:occupations})
  and (\ref{nm}), and  classical Monte Carlo calculations for a $19$-particle
   system (line).

4. (a) The number of holes that minimizes the energy of the Girvin wavefunction
 (\ref{eq:girvin}) for two different sizes of step potentials.
The straight lines correspond to $N_h\propto N/2$,
leading to an N-independent edge size.
 The inset shows the tunneling amplitude, as estimated from
 the occupation of the relevant state, as a function of $N$, on a log-log plot.
 The suppression of at least half of the peaks is evident and agrees very well
 with the theoretically predicted $1/N$ dependence \cite{jari}.
 (b) The density profile of the ground states of $N=30$ electrons for $3$
different
  potential slopes. The existence of a $\nu=1$ region is evident for the
highest
 slope, while for the other two slopes an incompressible $\nu=1/3$ strip is
formed
 along the edge.


\begin{references}
\bibitem[1]{wen} For a review see X.-G. Wen, {\sl Int. J. Mod. Phys.}
 {\bf 6}, 1711 (1992). {\sl ibid} {\bf 8},  457 (1994).
\bibitem[2]{leo}  L. P. Kouwenhoven et al., {\sl
   Phys. Rev. Lett.} {\bf 64},  685 (1990).
\bibitem[3]{simmons} J. A. Simmons et al., {\sl Phys. Rev. Lett.} {\bf 63},
 1731 (1989); V. J. Goldman and B. Su,  {\sl Science} {\bf 267},  1010 (1995).
\bibitem[4]{paul} P. L. McEuen et al.,
 \PRL {\bf 66}, 1926  (1991); A. A. M. Staring et al., \pr {\bf B46},  12869
 (1992); R. C. Ashoori et al., \prl {\bf 71},  613 (1993); J. P. Bird
 et al., \pr {\bf B 50},  14983 (1994);
N. C. Van der Vaart et al., \prl {\bf 73},  320 (1994).
\bibitem[5]{mac1} M.D. Johnson and A. MacDonald, \PRL {\bf 67}, 2060 (1991).
\bibitem[6]{jari}J. M. Kinaret et al.,
{\sl Phys. Rev.} {\bf B45}, 9489 (1992); {\sl ibid} {\bf B46},
681 (1992);
\bibitem[7]{meirav}U. Meirav, M. A. Kastner and S. J. Wind, \PRL {\bf65},
771 (1990);
\bibitem[8]{macdonald} A. H. MacDonald, \PRL {\bf 64}, 220 (1990);
\bibitem[9]{girvin}S.M. Girvin, {\sl Phys.\ Rev.\ }{\bf B 29}, 6012 (1984).
\bibitem[10]{laughlin} R.B. Laughlin, \prl {\bf 50}, 1395 (1983).

\bibitem[11]{yoshi} D. Yoshioka, {\sl J. Phys. Soc. Japan} {\bf 62},
 839 (1993); M. Greiter, {\sl Phys. Lett.} {\bf  B336},  48 (1994).
\bibitem[12]{beenfrac} A. M. Chang, {\sl Solid State Comm.} {\bf 74}, 871
(1990);
C. W. J. Beenakker, \PRL {\bf 64}, 216 (1990).
\bibitem[13]{chklovskii} D. B. Chklovskii, B. I. Shklovskii and
L. I. Glazman, \pr {\bf B46}, 4026 (1992).
\bibitem[14]{more} In fact, as the potential becomes softer,
 more incompressible states are expected to nucleate along
the edge of the sample. Such very smooth potentials are not dealt with
in this work.
\bibitem[15]{myfqh} Y. Meir,  \prl {\bf 72},  2624 (1994).
\bibitem[16]{brey} L. Brey,  \pr {\bf B50},  11861 (1994); D. B. Chklovskii
{\sl ibid} {\bf 51},  9895 (1995).
\bibitem[17]{difference} Note that there is a factor of $N$ difference
between the definition of the two-particle correlation function here and
that of Ref.\cite{myfqh}.
\bibitem[18]{expand} G. V. Dunne, {\sl Int. J. Mod. Phys.} {\bf B7}, 4783
(1993).
\bibitem[19]{metro}M. Metropolis et al.,
{\sl J. Chem. Phys.} {\bf 21}, 1087 (1953).
\bibitem[20]{datta} The one-particle distribution function for the $\nu=1/3$
state has been calculated numerically also by classical molecular dynamics
[N. Datta and R. Ferrari, preprint].
\bibitem[21]{mac3} S. Mitra and A. H. MacDonald, \pr {\bf B 48}, 2005 (1993).
\bibitem[22]{frank}U. Meirav, Ph.D. thesis, M.I.T;
 A. Kumar, S. E. Laux and F. Stern, \pr {\bf B 42}, 5166 (1990).
\bibitem[23]{fisher} C. L. Kane,  M. P. A. Fisher and J. Polchinski,
\prl {\bf 72},  4129 (1994); C. L. Kane and  M. P. A. Fisher,
\pr {\bf B 51},  13449 (1995).
\bibitem[24]{eisenstein} J. P. Eisenstein et al., \pr {\bf B 41},  7910
(1990); I. F. Herbut, \pr {\bf B 46},  15582 (1992).




\end{references}
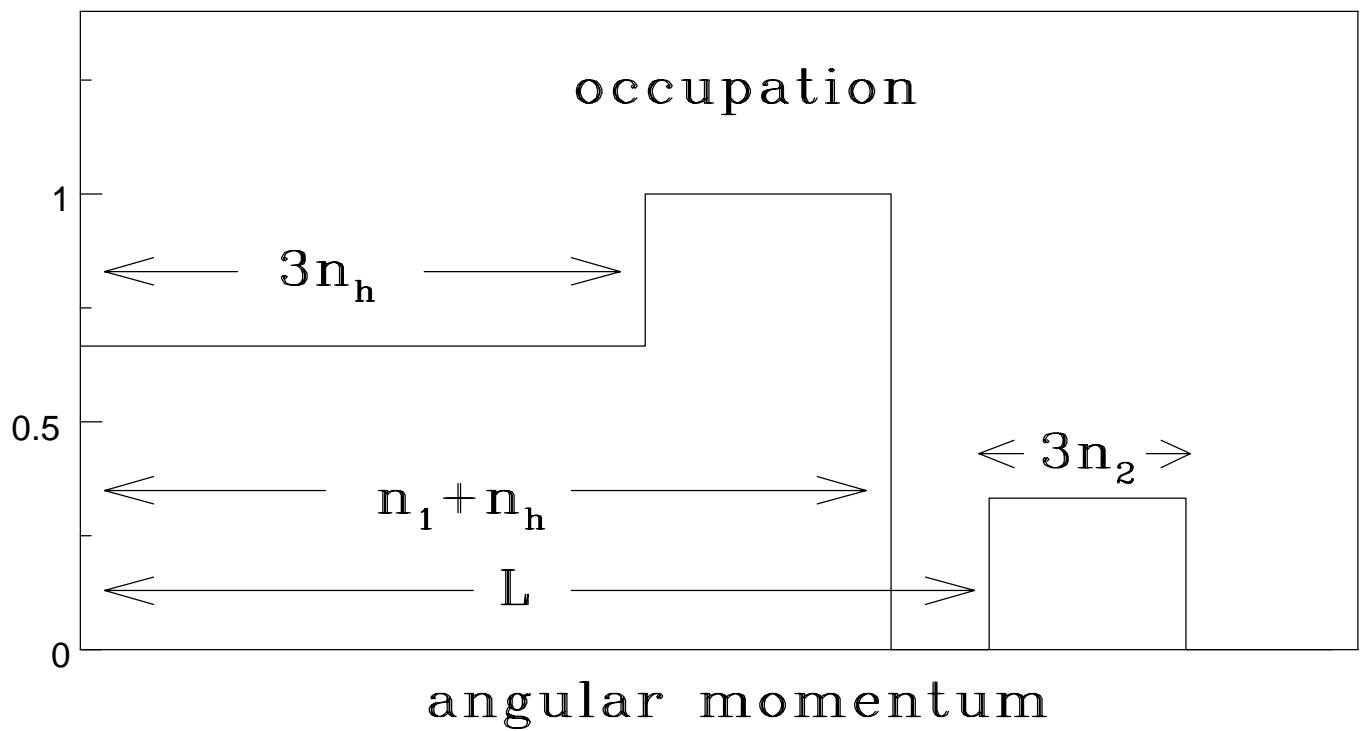
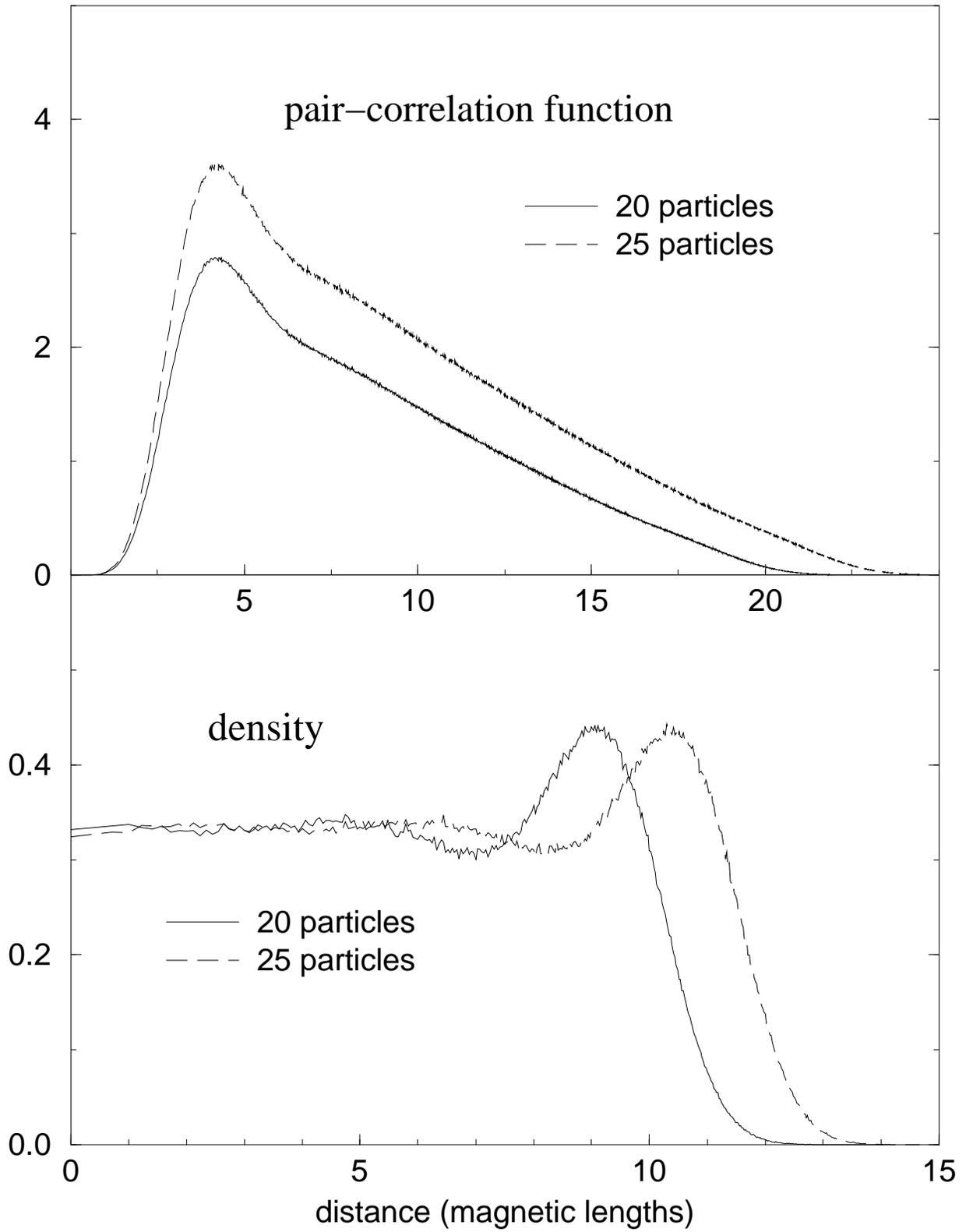
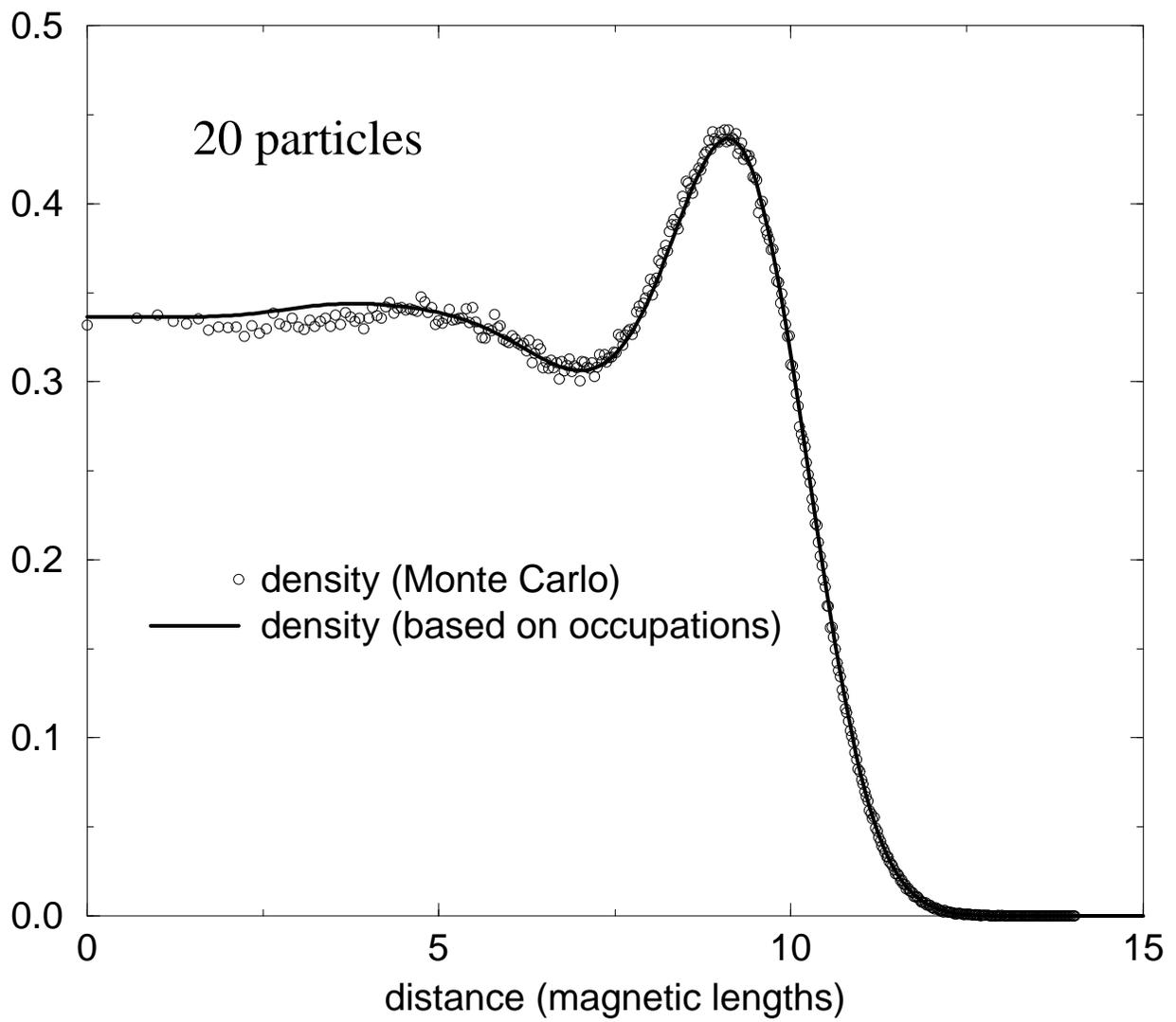
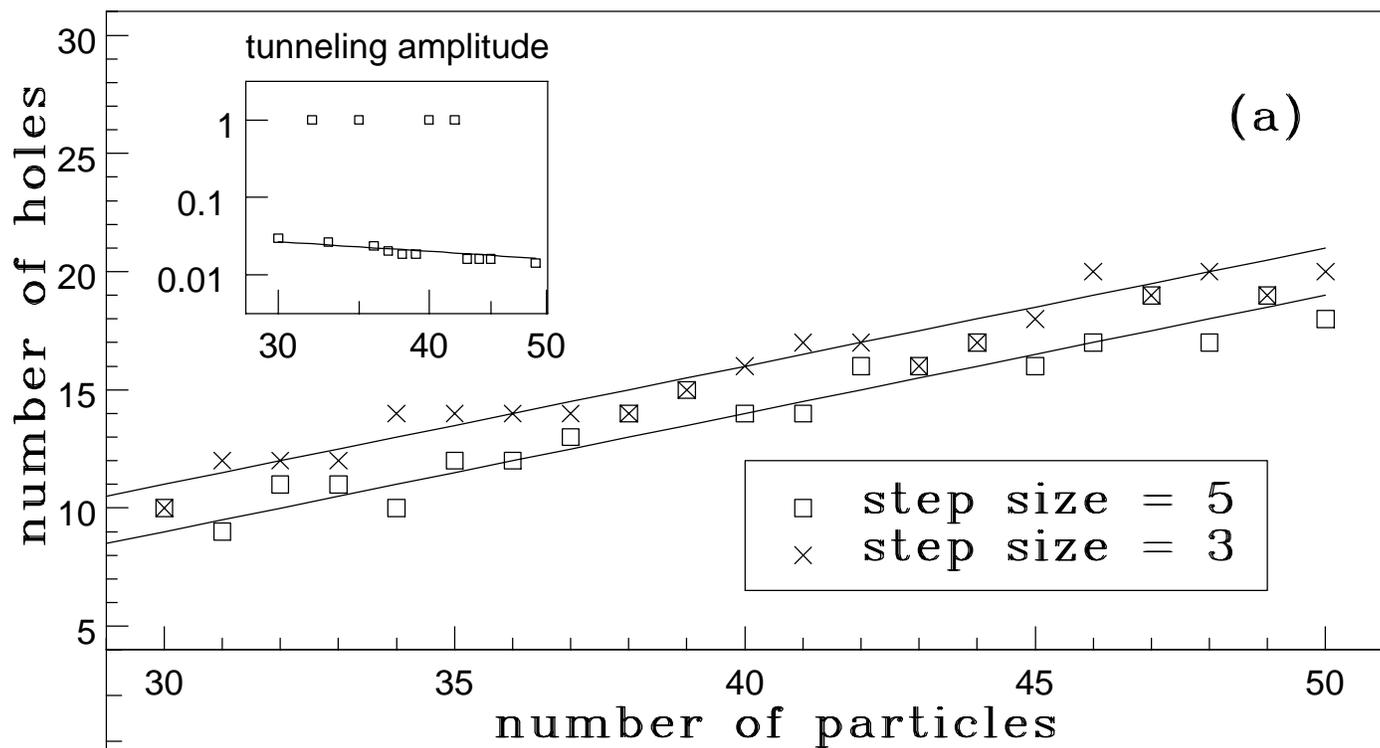
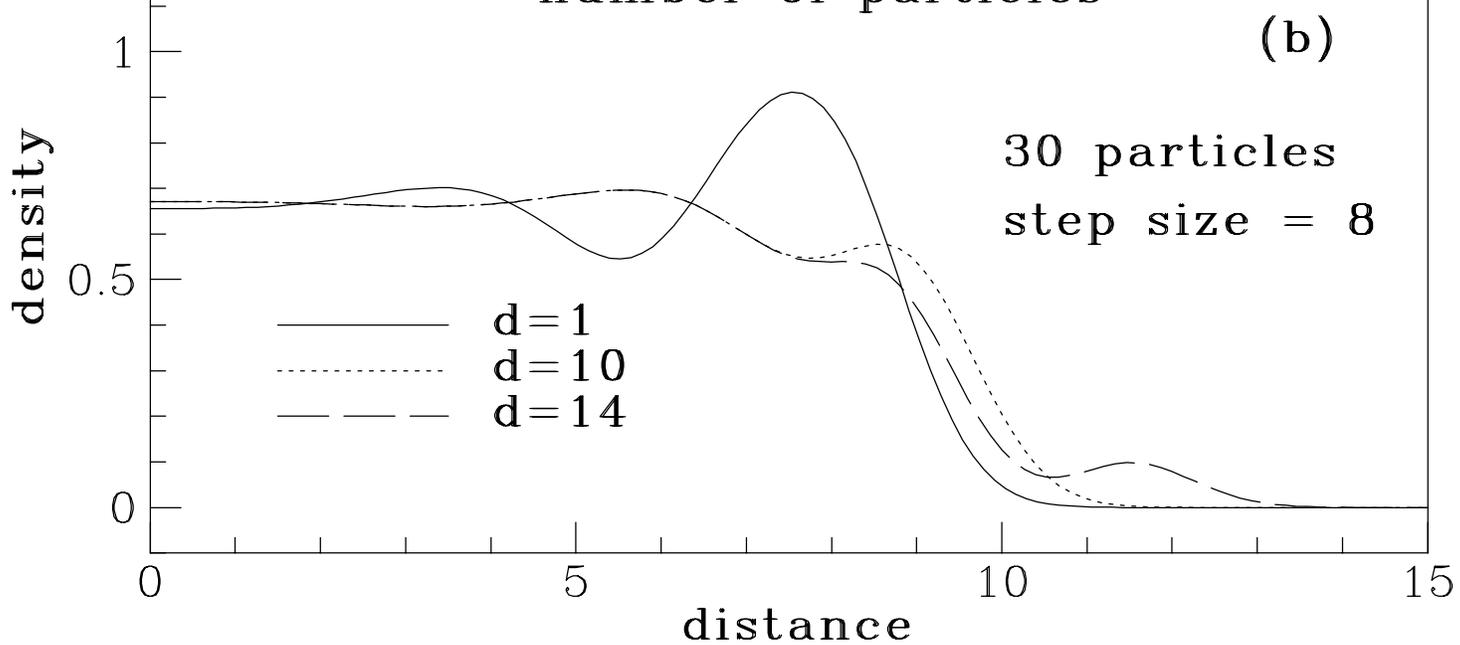
\end{document}